\documentclass{aip-cp}
\usepackage{cite}
\usepackage{rotating}
\usepackage{graphicx}
\usepackage{hyperref}
\usepackage{color}
\usepackage{epstopdf}
\usepackage{cleveref}
\usepackage[svgnames]{xcolor}
\usepackage{enumerate}
\usepackage{braket}
\hypersetup{hidelinks,colorlinks=true,allcolors=DarkBlue}
\usepackage{indentfirst}
\usepackage{amsfonts,amssymb}
\usepackage{subfigure}
\usepackage{caption}

\begin{document}

\title{Superfluidity of Identical Fermions in an Optical Lattice: Atoms and Polar Molecules}

\author[aff1]{A.K. Fedorov}\corresp[cor1]{Corresponding author: akf@rqc.ru (A.K.F.)}
\author[aff2,aff1]{V.I. Yudson}
\author[aff1,aff3,aff4,aff5,aff6,aff7]{G.V. Shlyapnikov}

\affil[aff1]{Russian Quantum Center, Skolkovo, Moscow 143025, Russia}
\affil[aff2]{Laboratory for Condensed Matter Physics, National Research University Higher School of Economics, Moscow 101000, Russia}
\affil[aff3]{SPEC, CEA, CNRS, Universit\'e Paris-Saclay, CEA Saclay, Gif sur Yvette 91191, France}
\affil[aff4]{LPTMS, CNRS, Univ. Paris-Sud, Universit\'e Paris-Saclay, Orsay 91405, France}
\affil[aff5]{Van der Waals-Zeeman Institute, Institute of Physics, University of Amsterdam, Science Park 904, 1098 XH Amsterdam, The Netherlands}
\affil[aff6]{Wuhan Institute of Physics and Mathematics, Chinese Academy of Sciences, 430071 Wuhan, China}
\affil[aff7]{Russian Quantum Center, National University of Science and Technology MISIS, Moscow 119049, Russia}

\maketitle

\begin{abstract}
In this work we discuss the emergence of $p$-wave superfluids of identical fermions in 2D lattices. 
The optical lattice potential manifests itself in an interplay between an increase in the density of states on the Fermi surface 
and the modification of the fermion-fermion interaction (scattering) amplitude. 
The density of states is enhanced due to an increase of the effective mass of atoms. 
In deep lattices, for short-range interacting atoms 
the scattering amplitude is strongly reduced compared to free space due to a small overlap of wavefunctions of fermions sitting in the neighboring lattice sites, 
which suppresses the $p$-wave superfluidity. 
However, we show that for a moderate lattice depth there is still a possibility to create atomic $p$-wave superfluids with sizable transition temperatures.
The situation is drastically different for fermionic polar molecules.
Being dressed with a microwave field, they acquire a dipole-dipole attractive tail in the interaction potential. 
Then, due to a long-range character of the dipole-dipole interaction, the effect of the suppression of the scattering amplitude in 2D lattices is absent.
This leads to the emergence of a stable topological $p_x+ip_y$ superfluid of identical microwave-dressed polar molecules. 
\end{abstract}

\section{Introduction}

The creation of $p_x+ip_y$ atomic or molecular topological superfluids in 2D optical lattices can be a promising path for quantum information processing~\cite{Nayak2008,Nayak2015},
since addressing qubits in the lattice should be much easier than in the gas phase.
For short-range interacting atomic fermions, 
the effect of the lattice potential on the formation of 
a superfluid phase of atomic fermions has been actively discussed~\cite{Lukin2002,LukinDemler2005,Iskin2005,Ketterle2006,Zoller2009,Lewenstein2010,Buchler2014,Liu2014}.
In particular, for the $s$-wave pairing of spin-1/2 fermions an increase in the depth of the optical potential results in a stronger atom localization and hence in increasing the on-site interaction.
At the same time, the tunneling becomes weaker.
The combined effect of these two factors is a strong increase in the critical temperature~\cite{Lukin2002,LukinDemler2005,Iskin2005}.
This has been observed in the MIT experiment~\cite{Ketterle2006}.
For the lattice filling somewhat smaller than unity, the physical picture can be rephrased as follows.
An increase in the lattice depth increases an effective mass of atoms and, hence, makes the density of states (DOS) larger.
The effective fermion-fermion scattering amplitude is also increasing.
The critical temperature in the BCS approach is $T_c\propto\exp\left[-1/\lambda_c\right]$,
where $\lambda_c$ is proportional to the product of the (modulus of) the scattering amplitude and the DOS on the Fermi surface.
Thus, an increase in the lattice potential increases the critical temperature.

On the contrary, for identical fermions in fairly deep lattices (tight-binding model) the fermion-fermion scattering amplitude is strongly reduced.
In the lowest band approach two fermions do not occupy the same lattice site,
and the amplitude is proportional to a very small overlap of the wavefunctions of fermions sitting in the neighboring sites.
This suppresses the $p$-wave superfluid pairing for fairly small filling factors in deep lattices, which is consistent with numerical calculations of Ref.~\cite{Iskin2005}.
Nevertheless, there remains a question about an interplay between an increase of the DOS and the modification of the fermion-fermion scattering amplitude for moderate lattice depths.
However, in sinusoidal optical lattices single particle states are described by complicated Mathieu functions, which complicates the analysis.
Therefore, we study identical fermionic atoms in a 2D version of the Kronig-Penney model allowing a transparent physical picture for moderate lattice depths. 

For microwave-dressed polar molecules the long-range character of the acquired attractive dipole-dipole intermolecular interaction strongly changes the situation. 
The interaction amplitude is not suppressed even in deep lattices, and a collisionally stable $p_x+ip_y$ superfluid may emerge. 

The work is organized as follows. 
First, we discuss superfluidity of identical atomic fermions in a 2D lattice. 
We describe a general approach for studying superfluidity of 2D lattice fermions.
We then show how the ordinary tight-binding optical lattice promotes the $s$-wave superfluidity of spin-1/2 fermionic atoms
and suppresses the $p$-wave superfluidity of spinless fermions.
Second, we develop a theory of $p$-wave superfluidity of spinless fermions in the 2D Kronig-Penney lattice and discuss inelastic decay processes.
We discuss $p$-wave superfluidity of identical microwave-dressed polar molecules 
and show that the long-range character of the dipole-dipole attraction leads to similar results regarding the critical temperature as in free space.
Finally, we summarize the main results of our work. 
The results of this work are based on Refs.~\cite{Fedorov2016,Fedorov2017}.

\section[Superfluidity of identical fermionic atoms]{Superfluidity of identical fermionic atoms in a 2D lattice}\label{sec:4atoms}

\subsubsection{General relations and qualitative arguments}

Let us first present a general framework for the investigation of superfluid pairing of weakly interacting lattice fermions.
We will do this for 2D identical (spinless) fermions, having in mind that the approach for spin-1/2 fermions is very similar.
The Hamiltonian of the system is 
\begin{equation}
	\hat{\mathcal{H}}=\hat{H}_{0} +\hat{H}_{\mathrm{int}},
\end{equation}
and the single particle part is given by (hereinafter in this Section we put $\hbar=1$ and set the normalization volume (surface) equal to unity):
\begin{equation}\label{hamiltonian0}
	\hat{H}_{0}={\int}d^2{\bf r}\,\hat\psi^{\dag}({\bf r}\,)\left[-\frac{\mathbf{\nabla}^2}{2m} + U({\bf r})-\mu\right]\hat\psi({\bf r}),
\end{equation}
with $\mu$ being the chemical potential,
$m$ the particle mass,
$U({\bf r})$ the 2D periodic lattice potential,
and $\hat\psi({\bf r})$ the fermionic field operator.
The term $\hat H_{\mathrm{int}}$ describes the interaction between particles:
\begin{equation}\label{hamiltonianint}
	\hat{H}_{\mathrm{int}}=\frac{1}{2}\int{d^2 r d^2 r'\,\hat\psi^{\dag}({\bf r})\hat\psi^{\dag}({\bf r}')V({\bf r}-{\bf r}')\hat\psi({\bf r}')\hat\psi({\bf r})},
\end{equation}
where $V({\bf r}-{\bf r}')$ is the potential of interparticle interaction of radius $r_0$.

In the absence of interactions,
fermions in the periodic potential $U({\bf r})$ fill single particle energy levels $\varepsilon_{\nu}({\bf k})$ determined by the Schr\"{o}dinger equation:
\begin{equation}\label{Single-particle}
	\left[-\frac{\mathbf{\nabla}^2}{2m}+U({\bf r}) \right]\chi_{\nu {\bf k}}({\bf r})=\varepsilon_{\nu}({\bf k})\chi_{\nu {\bf k}}({\bf r}).
\end{equation}
Here $\nu = 0,1,2, \ldots$ numerates energy bands, the wave vector ${\bf k}=\{k_x,k_y\}$ takes values within the Brillouin zone: 
\begin{equation}
	\{-\pi/b<k_i<\pi/b; i=x,y\},
\end{equation}
and $b$ is the lattice period.
The eigenfunctions $\chi_{\nu {\bf k}}({\bf r})$ obey the periodicity condition
\begin{equation}\label{chi-period}
	\chi_{\nu {\bf k}}({\bf r} + {\bf R}_n) =
	\chi_{\nu {\bf k}}({\bf r})\exp{[i{\bf k}{\bf R}_n]},
\end{equation}
where $n = (n_x, n_y)$ is the index of the lattice site, with integer $n_x, n_y$.
In the described Bloch basis the field operator reads:
\begin{equation}\label{Psi-expansion}
	\hat\psi({\bf r}){=}\sum\nolimits_{\nu, {\bf k}} \hat a_{\nu{\bf k}}\chi_{\nu {\bf k}}({\bf r}),
\end{equation}
with $\hat a_{\nu{\bf k}}$ being the annihilation operator of fermions with quasimomentum ${\bf k}$ in the energy band $\nu$.

We assume a dilute regime where the 2D density $n$ is such that $nb^2\lesssim1$, and
all fermions are in the lowest Brillouin zone (hereinafter we omit the corresponding index $\nu=0$). In the low momentum limit (small filling factor) that we consider, their Fermi
energy $E_F$ is small compared to the energy bandwidth $E_B$.
The lattice potential amplitude $U_0$ is assumed to be sufficiently large, so that
both $E_F$ and $E_B$ are smaller than the gap between the
first and second lattice bands.
The single particle dispersion relation then takes the form:
\begin{equation}\label{effdisp}
	E_k=\frac{k^2}{2m^*},
\end{equation}
where $m^*>m$ is the effective mass.

In 2D the transition of a Fermi gas from the normal to superfluid state is set by the Kosterlitz-Thouless mechanism. 
However, in the weakly interacting regime the Kosterlitz-Thouless transition  temperature is very close to $T_c$ calculated in the Bardeen-Cooper-Schrieffer (BCS) approach~\cite{Miyake1974}.
We then reduce the Hamiltonian given by Eqs. (\ref{hamiltonian0}) and (\ref{hamiltonianint}) to the standard BCS form:
\begin{equation}\label{hamiltonian-bcs}
	\hat{\mathcal{H}}_{\rm BCS}=
	\sum\nolimits_{{\bf k}}\left\{{(E_k-\mu)\hat{a}^{\dag}_{ {\bf k}}\hat{a}_{{\bf k}}^{}+\frac{1}{2}\left[\hat a^{\dag}_{{\bf k}}\hat a^{\dag}_{-{\bf k}}\Delta({\bf k}) + {\rm h.c.}\right]}\right\},
\end{equation}
where the momentum-space order parameter $\Delta({\bf k})$ is given by
\begin{equation}\label{Delta-definition}
	\Delta({\bf k})=\sum\nolimits_{{\bf k}\,'}V({\bf k},{\bf k}')\langle \hat a_{-{\bf k}'} \hat a_{{\bf k}'}\rangle,
	\qquad
	\Delta({\bf k})=-\Delta(-{\bf k}),
\end{equation}
with $V({\bf k},{\bf k}')$ being the matrix element of the interaction potential between the corresponding states.

The Hamiltonian (\ref{hamiltonian-bcs}) is then decomposed in a set of independent quadratic Hamiltonians
and the anomalous averages are determined by the standard BCS expressions:
\begin{eqnarray} \label{BCS-average}
	\langle{\hat a_{-{\bf k}} \hat a_{{\bf k}}}\rangle=-\Delta({\bf k})\mathcal{K}(k),
\end{eqnarray}
where $\mathcal{K}(k)=\tanh[\mathcal{E}(k)/2T]/2\mathcal{E}(k)$, 
and
\begin{eqnarray} \label{BCS energy}
	\mathcal{E}(k)=\sqrt{(E_k-\mu)^2 + |\Delta_{\nu}(\vec{k})|^2}
\end{eqnarray}
is the energy of excitation with quasimomentum ${\bf k}$.
From Eqs.~(\ref{Delta-definition}) and (\ref{BCS-average}) we have an equation
for $\Delta({\bf k})$ (gap equation):
\begin{equation}\label{Delta-equation}
	\Delta({\bf k})=-\sum\nolimits_{{\bf k}'}V({\bf k},{\bf k}')\mathcal{K}(k')\Delta({\bf k}').
\end{equation}
Eq.~(\ref{Delta-equation}) can be expressed \cite{Shlyapnikov2009,Shlyapnikov2011} 
in terms of the effective off-shell scattering amplitude $f({\bf k}',{\bf k})$ of a fermion pair with momenta ${\bf k}$ and $-{\bf k}$
defined as
\begin{equation}\label{scattering}
	f({\bf k}',{\bf k})=\int d^2r_{1} d^2r_{2}\,\Phi^{(0)*}_{{\bf k}'}({\bf r}_1,{\bf r}_2)V({\bf r}_1-{\bf r}_2)\Phi_{{\bf k}}({\bf r}_1,{\bf r}_2).
\end{equation}
Here
\begin{eqnarray} \label{Phi-0}
	\Phi^{(0)}_{{\bf k}}({\bf r}_1, {\bf r}_2)=\chi_{{\bf k}}({\bf r}_1)\chi_{-{\bf k}}({\bf r}_2),
\end{eqnarray}
is the wavefunction of a pair of non-interacting fermions with quasimomenta ${\bf k}$ and $-{\bf k}$.
The quantity $\Phi_{{\bf k}}({\bf r}_1,{\bf r}_2)$ is the true (i.e., accounting for the interaction) wavefunction,
which develops from the incident wavefunction $\Phi^{(0)}_{{\bf k}}({\bf r}_1,{\bf r}_2)$ of a free pair.
The wavefunction $\Phi_{{\bf k}}({\bf r}_1,{\bf r}_2)$ satisfies the Schr\"{o}dinger equation 
\begin{equation}
[\hat{H}_{12}-2E_k]\Phi_{{\bf k}}({\bf r}_1,{\bf r}_2)=0, 
\end{equation}
with the two-particle Hamiltonian:
\begin{eqnarray} \label{Phi-equation}
	\hat{H}_{12}=-\frac{\mathbf{\nabla}^2_1+\mathbf{\nabla}^2_2}{2m}+U({\bf r}_1)+U({\bf r}_2)+V({\bf r}_1{-}{\bf r}_2).
\end{eqnarray}
The renormalized gap equation for the function $\Delta({\bf k})$ then takes the form similar to that in free space (see Refs.~\cite{Shlyapnikov2009,Shlyapnikov2011} and references therein):
\begin{equation}\label{eq:4gap}
	\Delta({\bf k})=\int\frac{d^2k'}{(2\pi)^2}f({\bf k}',{\bf k})\Delta({\bf k}')\left\{\mathcal{K}(k')-\frac{1}{2(E_{k'}-E_k)}\right\}.
\end{equation}
In the weakly interacting regime the chemical potential coincides with the Fermi energy $E_F= k_F^2/2m^*$, where $k_F=\sqrt{4\pi{n}}$ is the Fermi momentum.
Note that here we omit a correction to the bare interparticle interaction due to polarization of the medium by colliding particles~\cite{Gor'kov1961}.

We will see below that the scattering amplitude and the corresponding critical temperature of the superfluid transition 
of lattice fermions depend drastically on the presence or absence of spin and on the pairing angular momentum.
Before analyzing various regimes, we discuss the situation in general.

The efficiency of superfluid pairing first of all depends on the symmetry of the order parameter.
For the pairing with orbital angular momentum $l$ we have 
\begin{equation}
	\Delta({\bf k})\rightarrow\Delta_l(k)\exp\left[il\phi_{\bf k}\right],
\end{equation}
where $\phi_{\bf k}$ is the angle of the vector ${\bf k}$ with respect to the quantization axis.
Integrating Eq.~(\ref{eq:4gap}) over $\phi_{\bf k}$ and $\phi_{{\bf k}'}$ we obtain the same equation in which $\Delta({\bf k})$ and $\Delta({\bf k}')$ are replaced with $\Delta_l(k)$ and $\Delta_l(k')$,
and $f({\bf k}',{\bf k})$ is replaced with its $l$-wave part
\begin{equation}\label{fl}
	f_l(k',k)=\int \frac{d\phi_{\bf k}d\phi_{{\bf k}'}}{(2\pi)^2}f({\bf k}',{\bf k})\exp\left[il\phi_{\bf k}-il\phi_{{\bf k}'}\right].
\end{equation}
Alternatively, we can write
\begin{equation}\label{flalt}
	f_l(k',k)=\int d^2r_1d^2r_2\Phi^{(0)*}_{lk'}({\bf r}_1,{\bf r}_2)V(|{\bf r}_1-{\bf r}_2|)\Phi_{lk}({\bf r}_1,{\bf r}_2).
\end{equation}
where the $l$-wave parts of the wavefunctions, $\Phi^{(0)}_{lk'}$ and $\Phi_{lk}$, are given by
\begin{eqnarray}
	&&\Phi^{(0)}_{lk'}({\bf r}_1,{\bf r}_2)=\int\frac{d\phi_{{\bf k}'}}{2\pi}\Phi^{(0)}_{{\bf k}'}({\bf r}_1,{\bf r}_2)\exp\left[il\phi_{{\bf k}'}\right],
	\label{Phi0l} \\
	&&\Phi_{lk}({\bf r}_1,{\bf r}_2)=\int\frac{d\phi_{{\bf k}}}{2\pi}\Phi_{{\bf k}}({\bf r}_1,{\bf r}_2)\exp\left[il\phi_{{\bf k}}\right].
	\label{Phil}
\end{eqnarray}
As well as in free space (see Ref.~\cite{Shlyapnikov2009,Shlyapnikov2011}), we turn from $f_l(k',k)$ to the (real) function
\begin{equation}
	\tilde f_l(k',k)=f_l(k',k)\left[1-i\tan\delta(k)\right],
\end{equation}
where $\delta(k)$ is the scattering phase shift. 
This leads to the gap equation:
\begin{equation}\label{eq:4gapk}
	\Delta_l(k)=-P\int\frac{d^2k'}{(2\pi)^2}\tilde{f}_l(k',k)\Delta_l(k')\left\{\mathcal{K}(k')-\frac{1}{E_{k'}-E_k}\right\},
\end{equation}
where the symbol $P$ denotes the principal value of the integral.

In order to estimate the critical temperature $T_c$, we first put $k=k_F$ and notice that the main contribution to the integral over $k'$ in Eq.~(\ref{eq:4gapk}) comes from $k'$ close to $k_F$.
At temperatures $T$ tending to the critical temperature $T_c$ from below, we put $\mathcal{E}(k')=|E_{k'}-E_F|$ in $\mathcal{K}(k')$.
Then for the pairing channel related to the interaction with orbital angular momentum $l$, we have the following estimate:
\begin{equation}\label{eq:Tc0}
	T_c\sim E_F\exp\left[-\frac{1}{\lambda_c}\right],
	\qquad
	\lambda_c=\rho(k_F)|f_l(k_F)|.
\end{equation}
The quantity $\rho(k_F){=}m^*/2\pi$ is the effective density of states on the Fermi surface,
and $f_l(k_F)$ is the on-shell $l$-wave scattering amplitude of lattice fermions.
The derivation for spin-1/2 fermions with attractive intercomponent interaction leads to the same gap equations (\ref{eq:4gap}), (\ref{eq:4gapk}) and estimate (\ref{eq:Tc0}) in which
\begin{equation}
	\Delta({\bf k})=\sum_{{\bf k}'}V({\bf k},{\bf k}')\langle\hat a_{\downarrow -{\bf k}'}\hat a_{\uparrow {\bf k}'}\rangle
\end{equation}
and $f({\bf k}',{\bf k})$, $f_l(k',k)$ are the amplitudes of the intercomponent interaction.

Eq.~(\ref{eq:Tc0}) shows that compared to free space we have an additional pre-exponential factor $m/m^*<1$. 
Assuming that the lattice amplitude $f_l(k_F)$ and the free-space amplitude $f_l^0(k_F)$ are related to each other as
\begin{equation}\label{f-lat}
	f_{l}(k_F)=\mathcal{R}_{l}f_l^0(k_F),
\end{equation}
we see that the exponential factor $\lambda_c$ in Eq.~(\ref{eq:Tc0}) becomes
\begin{equation}\label{eq:lambda}
	\lambda_{c}=\mathcal{R}_{l}\frac{m^*}{m}\lambda^0_c,
\end{equation}
where $1/\lambda^0_c$ is the BCS exponent in free space.
Below we compare $T_c$ in various lattice setups with the critical temperature in free space.

\subsubsection{Short-range interacting atomic fermions in a deep 2D lattice}\label{sec:4cosine}

We start with the analysis of superfluid pairing in deep 2D lattices. As an example, we consider a quadratic lattice with the lattice potential of the form:
\begin{eqnarray}\label{eq:U}
	U({\bf r})=U_0\left[\cos{\left(\frac{2\pi}{b}x\right)}+\cos{\left(\frac{2\pi}{b}y\right)}\right].
\end{eqnarray}
For sufficiently deep lattices, the single particle wavefunction has the Wannier form:
\begin{equation}\label{eq:wavefunction}
	\chi_{{\bf k}}({\bf r})=\frac{1}{\sqrt{\mathcal{N}}}\sum_j\phi_0({\bf r}-{\bf R}_j)\exp[i{\bf k}{\bf R}_j],
\end{equation}
where the ground state wavefunction in the lattice cell has an extention $\xi_0$ and is given by
\begin{equation}\label{eq:wavefunctionground}
	\phi_0({\bf r})=\frac{1}{\sqrt{\pi}\xi_0}\exp\left[-\frac{r^2}{2\xi_0^2}\right].
\end{equation}

Using a general formula for the effective mass from Ref.~\cite{LL9},
for a deep potential of the form (\ref{eq:U}) one obtains:
\begin{equation}\label{eq:mass}
	\frac{m^*}{m}\simeq\pi\frac{\xi_0^2}{b^2}\exp\left[\frac{2}{\pi^2}\frac{b^2}{\xi_0^2}\right].
\end{equation}

We will consider fermionic atoms interacting with each other via a short-range potential $V({\bf r})$ of radius $r_0$ and assume the following hierarchy of length scales:
\begin{equation}\label{eq:hierarchy}
	r_0\ll {\xi_{0}}<{b}<{1/k_{F}}.
\end{equation}
We first discuss the $s$-wave pairing of spin-1/2 fermions with attractive intercomponent interaction ($l=0$).

Turning to Eq.~(\ref{flalt}) for $l=0$,
we notice that the main contribution to the $s$-wave scattering amplitude in the lattice comes from the interaction between spin-up and spin-down fermions sitting in one and the same lattice site.
The wavefunctions $\Phi^{(0)}_{0k'}$ and $\Phi_{0k}$ can be written as
\begin{eqnarray}
	&&\Phi_{0k'}^{(0)}({\bf r}_1,{\bf r}_2)=\chi_0({\bf r}_1)\chi_0({\bf r}_2), \\
	&&\Phi_{0k}({\bf r}_1,{\bf r}_2)=\chi_0({\bf r}_1)\chi_0({\bf r}_2)\zeta_0(|{\bf r}_1-{\bf r}_2|),
\end{eqnarray}
where the function $\zeta_0(|{\bf r}_1-{\bf r}_2|)$ is a solution of the Schr\"{o}dinger equation for the $s$-wave relative motion of two particles in free space at zero energy,
and it is tending to unity for interatomic separations greatly exceeding $r_0$.
We put $l=0$ in Eq.~(\ref{flalt}) and integrate over ${\bf r}={\bf r}_1-{\bf r}_2$ and ${\bf r}_+=({\bf r}_1+{\bf r}_2)/2$.
Then, owing to the inequality $r_0\ll\xi_0$, this equation is reduced to
\begin{equation}   \label{f0}
	f_0(k',k)=\int d^2r\,V(r)\zeta(r)\,\int d^2r_+|\chi_0(r_+)|^4.
\end{equation}
Recalling that in the low momentum limit the free space scattering amplitude is given by
\begin{equation}
	f_0^0=\int V(r)\zeta(r)d^2r
\end{equation}
and using Eq.~(\ref{eq:wavefunction}) for the function $\chi_0(r)$, we obtain for the ratio of the lattice to free space amplitude:
\begin{equation}\label{eq:RS}
	\mathcal{R}_{l=0}=\frac{1}{2\pi}\frac{b^2}{\xi_0^2}.
\end{equation}
Therefore, according to Eqs. (\ref{eq:lambda}) and (\ref{eq:mass}) the BCS exponent $\lambda_c^{-1}$ becomes smaller than in free space by the following factor:
\begin{equation}\label{eq:enhanceSshort}
	\mathcal{R}_{l=0}\,\frac{m^*}{m}\simeq\frac{1}{2}\exp\left[\frac{2}{\pi^2}\frac{b^2}{\xi_0^2}\right].
\end{equation}

For example, taking $b/\xi_0=4$ the BCS exponent $\lambda_c^{-1}$ decreases by a factor of $0.08$,
whereas the effective mass becomes higher by a factor of 5 compared to the bare mass $m$ (see Fig.~1).
Then, for $^6$Li atoms at density $10^8$ cm$^{-2}$ ($b\simeq 250$ nm, $k_Fb\simeq 0.5$) we have the Fermi energy $\sim 40$ nK.
Assuming that the free space BCS exponent is about 30 and the related critical temperature is practically zero, in the lattice we obtain $T_c\sim 3$ nK.
We thus see that the lattice setup may strongly promote the $s$-wave superfluidity of spin-1/2 fermions.

\begin{figure}[h]
\begin{minipage}[h]{1\linewidth}
\centerline{\includegraphics[width=0.5\columnwidth]{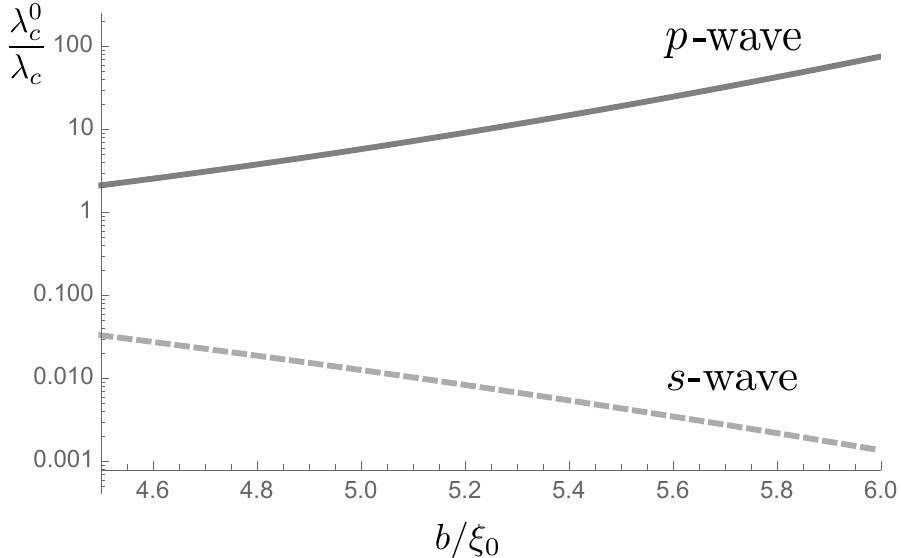}}
\centerline{\mbox{\textbf{Fig.~1}. The ratio of the BCS exponent in the tight--binding sinusoidal lattice to the BCS exponent in free space,}}
\centerline{$\lambda_c^0/\lambda_c$, at the same density and short-range coupling strength.}
\end{minipage}
\end{figure}

The situation with $p$-wave superfluidity of identical fermions is drastically different.
In the single band approximation (tight binding model) two such fermions can not occupy one and the same lattice site.
This is clearly seen using the functions $\chi_{\bf k}({\bf r}_1)$ and $\chi_{-{\bf k}}({\bf r}_2)$ from Eq.~(\ref{eq:wavefunction}) at the same ${\bf R}_j$,
so that the wavefunction $\Phi^{(0)}_{{\bf k}'}({\bf r}_1,{\bf r}_2)$  becomes independent of ${\bf k}'$.
Therefore, the $p$-wave part of this wavefunction $\Phi^{(0)}_{1k'}$ and the $p$-wave scattering amplitude $f_1(k',k)$ following from Eqs.~(\ref{Phi0l}) and (\ref{flalt}) at $l=1$ are equal to zero.

The main contribution to the interaction amplitude then comes from the overlap of the wavefunctions of fermions sitting in the neighbouring sites.
We then use Eqs. (\ref{eq:wavefunction}) and (\ref{eq:wavefunctionground}) and write:
\begin{equation}\label{Phi10} 
	\Phi^{(0)}_{{\bf k}'}({\bf r}_1,{\bf r}_2)=\chi_{{\bf k}'}({\bf r}_1)\chi_{-{\bf k}'}({\bf r}_2)
	=\frac{1}{\mathcal{N}\pi \xi_0^2}\sum_{i,j}\exp\left\{-\frac{({\bf r}_1-{\bf R}_i)^2}{2\xi_0^2}-\frac{({\bf r}_2-{\bf R}_j)^2}{2\xi_0^2}-i{\bf k}'{\bf b}_j\right\},  
\end{equation}
with ${\bf b}_j={\bf R}_j-{\bf R}_i$ and ${\bf R}_i,{\bf R}_j$ being the coordinates of the sites $i$ and $j$.  
For the short-range interaction between particles the main contribution to the scattering amplitude comes from distances ${\bf r}_1,{\bf r}_2$ 
that are very close to each other, and for given $i,j$ both coordinates should be close to $({\bf R}_j+{\bf R}_i)/2$. 
Therefore, Eq.~(\ref{Phi10}) is conveniently rewritten as 
\begin{equation}\label{Phi20}
	\Phi^{(0)}_{{\bf k}'}({\bf r}_1,{\bf r}_2)=\frac{1}{\mathcal{N}\pi\xi_0^2}\sum_{i,j}\exp\Big\{-i{\bf k}'{\bf b}_j-\frac{r_{+j}^2}{\xi_0^2}-
	\frac{r^2}{4\xi_0^2}-\frac{b^2}{4\xi_0^2}-\frac{{\bf r}{\bf b}_j}{2\xi_0^2}\Big\},    
\end{equation}
where ${\bf r}={\bf r}_1-{\bf r}_2$, ${\bf r}_{+j}={\bf r}_+-({\bf R}_i+{\bf R}_j/2$, ${\bf r}_+=({\bf r}_1+{\bf r}_2)/2$, 
and the summation is performed over the sites $j$ that are nearest neighbours of the site $i$. 
Assuming the conditions $k'b\ll 1$ and $r\sim r_0\ll \xi_0^2/b\ll\xi_0$, for the $p$-wave part of this wavefunction equation (\ref{Phi0l}) at $l=1$ gives:
\begin{equation}\label{Phi0p} 
	\Phi^{(0)}_{1k'}(r,r_+,\phi_{\bf r})=\frac{k'rb^2}{\mathcal{N}8\pi\xi_0^4}\sum_{i,j}\exp\left\{-\frac{r_{+j}^2}{\xi_0^2}-\frac{b^2}{4\xi_0^2}\right\}[\exp(i\phi_{\bf r})+\exp(-i\phi_{\bf r}+2i\phi_j)],   
\end{equation}
where $\phi_{\bf r}$ and $\phi_j$ are the angles of the vectors ${\bf r}$ and ${\bf b}$ with respect to the quantization axis. 
The $p$-wave part of the true relative-motion wavefunction $\Phi_{\bf k}({\bf r}_1,{\bf r}_2)$ under the same conditions is given by
\begin{equation}\label{Phip}
	\Phi_{1k}(r,r_+,\phi_{\bf r})=\frac{b^2}{\mathcal{N}4\pi\xi_0^4}\zeta_1(r)\sum_{i',j'}\exp\left\{-\frac{r_{+j'}^2}{\xi_0^2}-\frac{b^2}{4\xi_0^2}\right\}[\exp(i\phi_{\bf r})+\exp(-i\phi_{\bf r}+2i\phi_j)].
\end{equation}
The function $\zeta_1(r)$ is a solution of the Schr\"{o}dinger equation for the $p$-wave relative motion of two particles at energy tending to zero in free space. 
Sufficiently far from resonance, where the on-shell scattering amplitude satisfies the inequality $m|f_1(k)|\ll 1$, the function $\zeta_1(r)$ becomes $kr/2$ at distances $r\gg r_0$.

Looking at the product of the free and true relative-motion wavefunctions we notice that the main contribution to the scattering amplitude (\ref{flalt}) comes from the terms in which 
${\bf R}_i+{\bf R}_j={\bf R}_{i'}+{\bf R}_{j'}$, i.e. ${\bf r}_{+j}={\bf r}_{+j'}$. 
This is realized for $i=i'$, $j=j'$ or $i'=j$, $j'=i$.
Then, recalling that for $k'r_0\ll 1$ and $kr_0\ll 1$ the free space off-shell scattering amplitude is
\begin{equation}
	f_1^0(k',k)=\int V(r)(k'r/2)\zeta_1(r)d^2r,
\end{equation}
we first integrate each term of the sum over $i,j,i'j'$ in the product $\Phi^{(0)*}_{1k'}\Phi_{1k}$ over $d^2r$ and $d^2r_+$ in Eq.~(\ref{flalt}). 
After that we make a summation over the neighbouring sites $j$ and over the sites $i$ and take into account that $\mathcal{N}=1/b^2$. 
Eventually, this gives for the ratio of the lattice to free space $p$-wave amplitude:
\begin{equation}\label{Rl1}
	\mathcal{R}_{l=1}=\frac{1}{2\pi}\left(\frac{b}{\xi_0}\right)^6\exp\left[-\frac{b^2}{2\xi_0^2}\right].
\end{equation}
Thus, with the help of Eq.~(\ref{eq:mass}) the inverse BCS exponent in the lattice becomes:
\begin{equation}\label{lambdac1}
	\lambda_c=\mathcal{R}_{l=1}\frac{m^*}{m}\lambda_c^0=\frac{\lambda_c^0}{2}\left(\frac{b}{\xi_0}\right)^4\exp\left[-\frac{cb^2}{\xi_0^2}\right],
\end{equation}
where $c\simeq{0.3}$.

We now clearly see that the inverse BCS exponent $\lambda_c$ in the lattice is exponentially small compared to its value in free space.
In particular, already for $b/\xi_0=5$ the ratio $\lambda_c^0/\lambda_c$ it is about $6$, which practically suppresses $p$-wave superfluidity of identical fermions (see Fig.~1).
However, this ratio rapidly reduces with decreasing the ratio $b/\xi_0$ and becomes $\sim 1$ for $b/\xi_0=4$.
It is therefore interesting to analyze more carefully the case of moderate lattice depths.

\begin{figure}[h]
\begin{minipage}[h]{1\linewidth}
\centerline{\includegraphics[width=0.5\columnwidth]{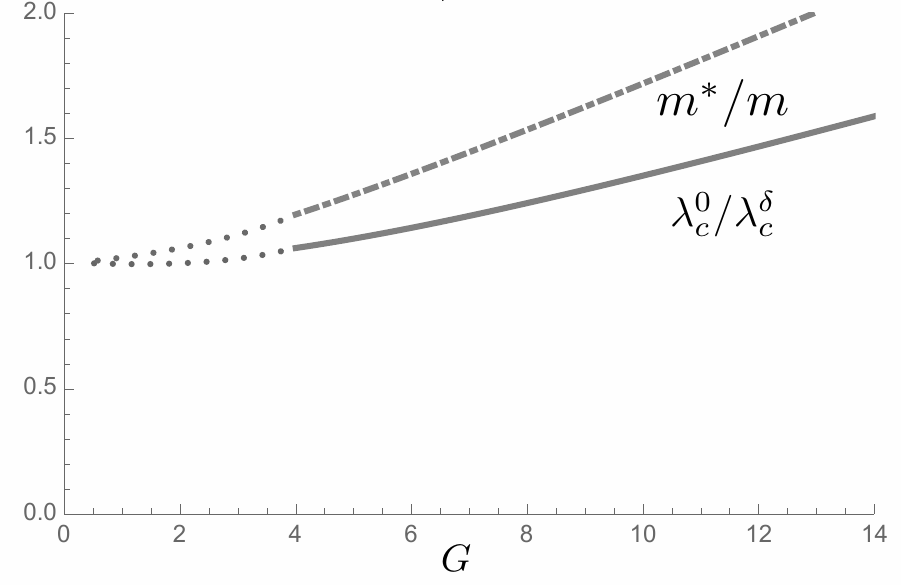}}
\centerline{\mbox{\newline \textbf{Fig.~2}. The ratio of the BCS exponent in the 2D $\delta$-functional Kronig-Penney lattice to the BCS exponent in free space,}}
\centerline{$\lambda_c^0/\lambda_c^{\delta}$, at the same density and short-range coupling strength. 
The solid blue curve shows $\lambda_c^0/\lambda_c^{\delta}$ as a function of the lattice depth $G$,}
\centerline{and the dashed red curve the effective mass $m^*/m$ versus $G$. The dotted parts of these curves show our expectation at $G\lesssim 1$, }
\centerline{where the single-band approximation used in our calculations does not work.}
\end{minipage}
\end{figure}

\subsubsection{Superfluid $p$-wave pairing in the 2D Kronig-Penney lattice}\label{sec:4KP}

We will do so using a 2D version of the Kronig-Penney model,
namely a superposition of two 1D Kronig-Penney lattices (in the $x$ and $y$ directions, respectively), with a $\delta$-functional form of potential barriers:
\begin{equation}\label{eq:UKP}
	U(x,y)=U_{0}b\sum_{j=-\infty}^{+\infty}\left[\delta(x-jb)+\delta(y-jb)\right].
\end{equation}
With the eigenfunctions being piecewise plane waves,
the 1D Kronig-Penney potential is used in ultracold atom theory (see, e.g. \cite{Wu2007,Calarco2014,BaranovZoller2016,Gorshkov2016}) to mimic sinusoidal potentials.
The model (\ref{eq:UKP}) catches the key physics and allows for transparent calculations.
The latter circumstance is a great advantage compared to sinusoidal lattices where single particle states are described by complicated Mathieu functions.
The considered model allows us to investigate two important questions.
The first question is about an interplay between an increase of the DOS and the modification of the fermion-fermion interaction for moderate lattice depths.
The second one is the stability of the system with respect to collisional losses.

Single-particle energies in the periodic potential (\ref{eq:UKP}) are represented as
\begin{equation}\label{eq:DKP}
	E_{\textbf k}=E(k_x)+E(k_y),
\end{equation}
where $E(k_{x,y})>0$ is the dispersion relation for the 1D Kronig--Penney model.
It follows from the equation (see, e.g., Ref.~\cite{LL9}):
\begin{equation}\label{eq:DKP}
	\cos(q{b})+G\frac{\sin(q{b})}{qb}=\cos(kb),
\end{equation}
where $q{=}\sqrt{2mE(k)}>0$, and $G=mU_0b^2$.
As well as in the previous section,
we consider a dilute regime where the filling factor is $\nu=nb^2{\lesssim}1$ and the fermions fill only a small energy interval near the bottom of the lowest Brillouin zone.
Then the energy counted from the bottom of the zone is given by Eq.~(\ref{effdisp}) and for the effective mass Eq.~(\ref{eq:DKP}) yields:
\begin{equation}\label{eq:mKP}	
	\frac{m^*}{m}\approx\frac{\tan(\eta/2)}{\eta}\left[{1+\frac{\sin\eta}{\eta}}\right],
\end{equation}
with $\eta$ being the smallest root of the equation:
\begin{equation}\label{eq:eta}
	\eta\tan(\eta/2)=G.
\end{equation}
Actually, $\eta=q_0b$ where $q_0$ follows from Eq.~(\ref{eq:DKP}) at $k=0$.

For $m^*\gg{m}$ we have $m^*/m=G/\pi^2$, which means that the quantity $G$ should be very large.
Then the width of the lowest Brillouin zone is $E_B=2/m^*b^2$ and it is much larger than the Fermi energy $E_F=k^2/2m^*$ for $k_Fb<0.5$.
The gap between the lowest and second zones is $E_G=3\pi^2/2mb^2$ and it greatly exceeds $E_B$ and $E_F$.
Note that even for $m^*\simeq1.3m$ ($G\simeq 5$) we have $E_G$ close to $4E_B$, and the ratio $E_F/E_B$ is significantly smaller than unity if $k_Fb<0.5$.
This justifies the single-band approximation and the use of the quadratic dispersion relation (\ref{effdisp}).

Single-particle wavefunctions $\chi_{\bf k}(\bf r)$ are of the form $\chi_{\bf k}({\bf r})=\chi_{k_x}(x)\chi_{k_y}(y)$, where
\begin{equation}\label{eq:chi-exact}
	\chi_{k_x}{(x)}{=}\frac{\sqrt{2}\sin\left(\eta/2\right)}{\sqrt{1+{\sin\eta}/{\eta}}}\sum_{j=-\infty}^{j=+\infty}{A_j(x)\exp{[ik_xjb]}}
	\left\{\frac{e^{iq{b}}e^{iq{(x-jb)}}}{e^{iq{b}}-e^{ik_x{b}}}-\frac{e^{-iq{b}}e^{-iq{(x-jb)}}}{e^{-iq{b}}-e^{ik_x{b}}}\right\}
\end{equation}
is the exact eigenfunction of the 1D Kronig-Penney model, with $A_j(x)=1$ for $(j-1)b<x<jb$ and zero otherwise.
The function $\chi_{k_y}(y)$ has a similar form.
For $k'b\ll 1$ and $kb\ll 1$ the $p$-wave parts of the wavefunctions, $\Phi^{(0)}_{1k'}$ and $\Phi_{1k}$, following from Eqs. (\ref{Phi-0}), (\ref{Phi0l}), and (\ref{Phil}) at $l=1$ turn out to be
\begin{eqnarray}
	&&\Phi^{(0)}_{1k'}=ik'r\frac{\eta\cot(\eta/2)}{[1+\sin\eta/\eta]^2}\sum_{j_x,j_y=-\infty}^{\infty}A_{j_x}(x_+)A_{j_y}(y_+) \label{Phi0KP}
	\left\{\cos\phi_{\bf r}\cos^2\left(q_0y_+-j_yb+\frac{b}{2}\right)+i\sin\phi_{\bf r}\cos^2\left(q_0x_+-j_xb+\frac{b}{2}\right)\right\};  \nonumber \\
	&&\Phi_{1k}=2i\zeta_1(r)\frac{\eta\cot(\eta/2)}{[1+\sin\eta/\eta]^2}\sum_{j_x,j_y=-\infty}^{\infty}A_{j_x}(x_+)A_{j_y}(y_+)\label{PhiKP}
	\left\{\cos\phi_{\bf r}\cos^2\left(q_0y_+-j_yb+\frac{b}{2}\right)+i\sin\phi_{\bf r}\cos^2\left(q_0x_+-j_xb+\frac{b}{2}\right)\right\}, \nonumber
\end{eqnarray}
where the function $\zeta_1(r)$ is defined after equation (\ref{Phip}).
For the ratio of the lattice to free space scattering amplitude we then obtain:
\begin{equation}\label{RlKP}
	\mathcal{R}_{l=1}=\frac{\eta^2\cot^2{\left({\eta}/{2}\right)}}{\left[1+{\sin\eta}/{\eta}\right]^{4}}\left[\frac{3}{2}+\frac{2\sin\eta}{\eta}+\frac{\sin{2\eta}}{4\eta}\right],
\end{equation}
and using Eq.~(\ref{eq:mKP}) the inverse BCS exponent in the lattice is expressed through the inverse BCS exponent in free space as
\begin{equation}\label{lambdacdelta}
	\lambda_{c}^{\delta}=\mathcal{R}_{l=1}\frac{m^*}{m}\lambda_{c}^{0}=
	\frac{\eta\cot(\eta/2)}{\left[1+{\sin\eta}/{\eta}\right]^{3}}\left[\frac{3}{2}+\frac{2\sin\eta}{\eta}+\frac{\sin{2\eta}}{4\eta}\right]\lambda_{c}^{0}.
\end{equation}

In the extreme limit of $G\gg1$ we have $\eta\simeq (\pi-2\pi/G)$, so that $\mathcal{R}_{l=1}\simeq\pi^4/G^2$ and $\lambda^{\delta}_c/\lambda_c^0\simeq\pi^2/G\ll{1}$.
We thus arrive at the same conclusion as in the previous section for sinusoidal lattices: in a very deep lattice the $p$-wave pairing of identical fermions is suppressed.
However, even for $G\simeq{20}$ the BCS exponent in the lattice exceeds the exponent in free space only by a factor of $1.7$ at the same density and short-range coupling strength 
(see Fig.~2). 
It is thus crucial to understand what happens with the rates of inelastic decay processes in the lattice setup.

A detailed derivation of the rates of two-body inelastic relaxation and three-body recombination is given in Ref.~\cite{Fedorov2017}.
We obtain that in the lattice the two-body inelastic relaxation is reduced by a factor of $\mathcal{F}_2$ compared to free space:
\begin{equation}\label{w2fin}
	W_2^{\rm lat}=\mathcal{F}_2(\eta)W_2^{\rm free}.
\end{equation}
The function $\mathcal{F}_2(\eta)$ is displayed in Fig.~3 versus the lattice depth $G$, 
which is related to $\eta$ by Eq.~(\ref{eq:eta}).

The relation between the three-body recombination decay rate in free space and the one in the 2D lattice reads:
\begin{equation} \label{w3fin}
	W_3^{\rm lat}=W_3^{\rm free}\mathcal{F}_3(\eta).
\end{equation}
The function $\mathcal{F}_3$ is shown in Fig.~3 versus the lattice depth $G$ related to $\eta$ by Eq.~(\ref{eq:eta}). 
The results of this Section indicate that both two-body and three-body inelastic collisions are significantly suppressed in the lattice setup even at moderate depths.

The obtained results indicate that there are possibilities to create the superfluid topological $p_x+ip_y$ phase of atomic lattice fermions. 
In deep lattices the $p$-wave superfluid pairing is suppressed 
and even for moderate lattice depths the BCS exponent is larger than in free space at the same density and short-range coupling strength.
However, the lattice setup significantly reduces the inelastic collisional losses, 
so that one can get closer to the $p$-wave Feshbach resonance and increase the interaction strength without inducing a rapid decay of the system. 

\begin{figure}[h!]
\begin{minipage}[h]{1\linewidth}
\centerline{\includegraphics[width=0.5\columnwidth]{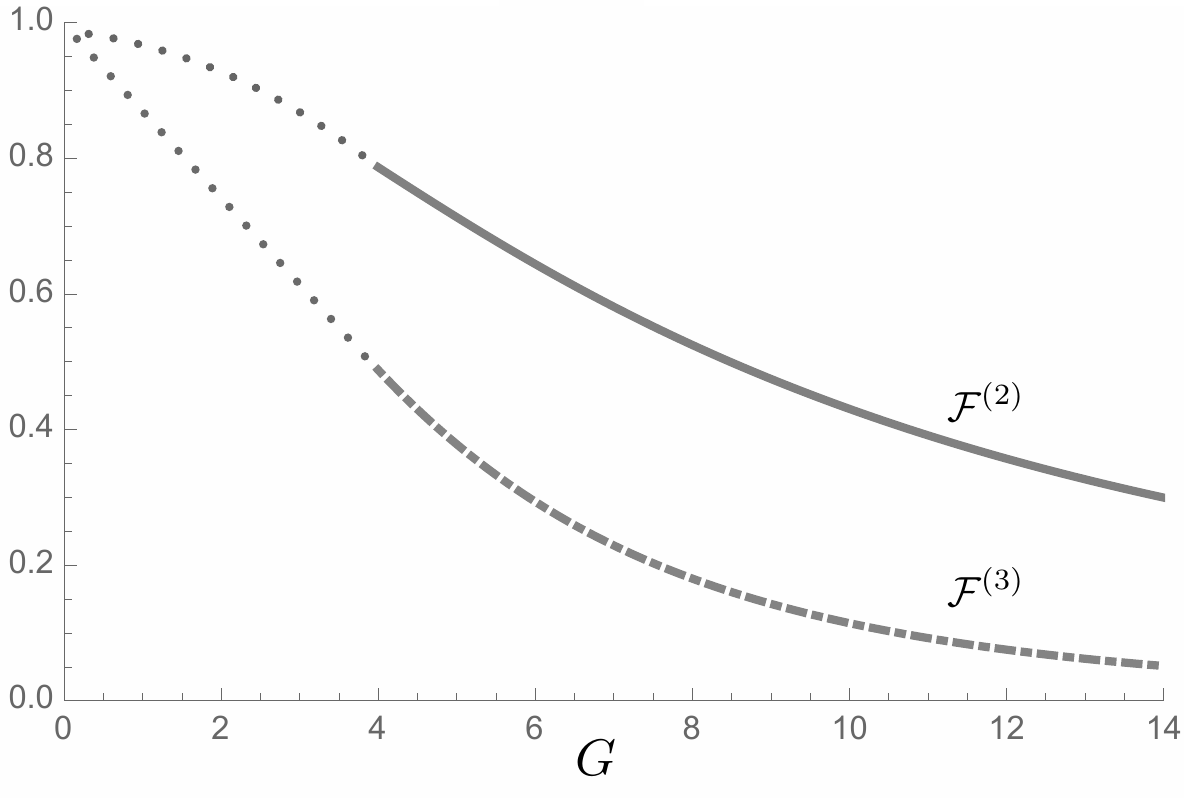}}
\centerline{\mbox{\textbf{Fig.~3}. Coefficients $\mathcal{F}^{(2)}$ and $\mathcal{F}^{(3)}$ as functions of the lattice depth $G$. The dotted parts of the curves show our}}
\centerline{expectation at $G\lesssim 1$, where the single-band approximation used in our calculations does not work.}
\end{minipage}
\end{figure}

For ultracold $^6$Li the $p$-wave resonance is observed for atoms in the lowest hyperfine state 
$(1/2,1/2)$~\cite{Salomon2004,Ketterle2005,Salomon2005,TicknorVale2008,UedaMukaiyama2008,Turlapov2010,Nakasuji2013}, 
and the only decay channel is three-body recombination. 
In the 2D Kronig-Penney lattice with the depth $G\simeq 12$ and the period $b\simeq 200$ nm ($m^*/m\simeq 2$ and $\mathcal{R}_{l=1} m^*/m\approx 0.7$), 
at $k_Fb\simeq 0.5$ the Fermi energy is close to 100 nK and the 2D density is about $0.5\times 10^8$ cm$^{-2}$. 
Slightly away from the Feshbach resonance (at the scattering volume $V_{sc}\simeq{8}\times{10^{-15}}$ cm$^3$) we are still in the weakly interacting regime, 
and the 3D recombination rate constant is $\alpha_{rec}^{3D}\sim 10^{-24}$ cm$^6$/s~\cite{Salomon2004}. 
Then, using Eq.~(\ref{eq:Tc0}) and the quasi2D scattering amplitude expressed through $V_{sc}$ and the tight confinement length $l_0=\sqrt{1/m\omega_0}$~\cite{Pricoupenko2008},
 for the confinement frequency $\omega_0{\simeq}100$ kHz we obtain the BCS critical temperature $T_c\simeq 5$ nK. 
 The 2D recombination rate constant is $\alpha_{rec}^{2D}\approx \mathcal{F}_3\alpha_{rec}^{3D}/\sqrt{3}\pi l_0^2$ 
 and with $\mathcal{F}_3\simeq 0.05$ at $G\simeq 12$ we arrive at the decay time $\tau_{rec}\sim 1/\alpha_{rec}^{2D}n^2$ approaching 1 second.
 
The $p$-wave Feshbach resonance for $^{40}$K occurs between atoms in the excited hyperfine state $(9/2,-7/2)$. 
Therefore, there is also a decay due to two-body relaxation. 
For the same parameters as in the discussed Li case ($G, V_{sc}, l_0, b, n$) we then have the Fermi energy $E_F\simeq 20$ nK, and the BCS transition temperature approaches 1 nK. 
Using experimental values for the relaxation and recombination rate constants in 3D~\cite{Jin2003} and retransforming them to the 2D lattice case, 
we obtain the relaxation and recombination times of the order of seconds.

\section[$P$-wave superfluids of molecules]{$P$-wave superfluids of fermionic polar molecules in a 2D lattice}\label{sec:4molecules}

\subsubsection{Scattering problem for microwave-dressed polar molecules in 2D lattices}\label{sec:4scattering}

We will consider identical fermionic polar molecules in a 2D lattice of period $b$. 
Being dressed with a microwave field, they acquire an attractive dipole-dipole tail in the interaction potential~\cite{BuchlerZoller2007,Gorshkov2008,Shlyapnikov2009,Shlyapnikov2011}:  
\begin{equation}\label{eq:4V}
	V(r)=-d^2/r^3.
\end{equation}
Here $d$ is an effective dipole moment, and we assume that Eq.~(\ref{eq:4V}) is valid at intermolecular distances $r\gtrsim{b}$. 

In the low momentum limit at a small filling factor the system of lattice polar molecules is equivalent to that of molecules with effective mass $m^{*}$ in free space.
We now demonstrate this explicitly by the calculation of the off-shell scattering amplitude $f({\bf k}',{\bf k})$.
For our problem the main part of the scattering amplitude can be obtained in the Born approximation~\cite{Shlyapnikov2009,Shlyapnikov2011}.  

In the lattice the scattering amplitude is, strictly speaking, the function of both incoming quasimomenta ${\bf q}_1,{\bf q}_2$ and outgoing quasimomenta ${\bf q}'_1,{\bf q}'_2$. 
However, in the low-momentum limit where ${qb}\ll{1}$, 
taking into account the momentum conservation law the amplitude becomes the function of only relative momenta ${\bf k}{=}({\bf q}_1-{\bf q}_2)/2$ and ${\bf k}'{=}({\bf q}'_1-{\bf q}'_2)/2$. 
For the off-shell scattering amplitude the first Born approximation gives:
\begin{equation}\label{Bf} 
	f({\bf k}',{\bf k})=S{\int\chi^*_{{\bf q}'_1}({\bf r}_1)\chi^*_{{\bf q}'_2}({\bf r}_2)V({\bf r}_1-{\bf r}_2)\chi_{{\bf q}_1}({\bf r}_1)\chi_{{\bf q}_2}({\bf r}_2)d^2r_1d^2r_2}
	=-\frac{d^2b^4}{S}\sum_{{{\bf r}_j},{{\bf r}'_j}}\frac{\exp[i({\bf q}_1-{\bf q}'_1){\bf r}_j+i({\bf q}_2{\bf q}'_2){\bf r}'_{j}]}{|{\bf r}_j-{\bf r}'_{j}|^3},
\end{equation}
where $V({\bf r}_1-{\bf r}_2)$ is given by Eq. (\ref{eq:4V}), and $S$ is the surface area.
The last line of Eq. (\ref{Bf}) is obtained assuming the tight-binding regime, where the single particle wavefunction is
\begin{equation}\label{psis}
	\chi_{{\bf q}}({\bf r})=\frac{1}{\sqrt{N}}\sum_j\Phi_0({\bf r}-{\bf r}_j)\exp\left[i{\bf q}{\bf r}_j\right].
\end{equation}
Here, the index $j$ labels the lattice sites located at the points ${\bf r}_j$, and $N=S/b^2$ is the total number of sites.
The particle wavefunction in a given site $j$ has extension $\xi_0$ and is expressed as 
\begin{equation}	
	\Phi_0({\bf r}-{\bf r}_j)=(1/\sqrt{\pi}\xi_0)\exp[-({\bf r}-{\bf r}_j)^2/2\xi_0^2].
\end{equation}
In the low-momentum limit we may replace the summation over $j$ and $j'$ by the integration over $d^2r_j$ and $d^2r'_{j}$ taking into account that $b^2\sum_j$ transforms into $\int d^2r_j$. 
This immediately yields
\begin{equation}\label{flm}
	f({\bf k}',{\bf k})=-d^2\int\exp[i({\bf k}-{\bf k}'){\bf r}]\frac{d^2r}{r^3},
\end{equation}
and the $p$-wave part of the scattering amplitude is obtained multiplying Eq. (\ref{flm}) by $\exp(-i\phi)$ and integrating over $d\phi/2\pi$, 
where $\phi$ is the angle between the vectors ${\bf k}$ and ${\bf k}'$. 
This is the same result as in free space (see, e.g., Refs.~\cite{Shlyapnikov2009,Shlyapnikov2011}). 
The on-shell amplitude ($k=k'$) can be written as 
\begin{equation}
	f(k)=-(8\hbar^2/3m^*)kr^*_{\rm eff}, 
\end{equation}
where $r^*_{\rm eff}=m^*d^2/\hbar^2$ is the effective dipole-dipole distance in the lattice. 
The applicability of the Born approximation assumes that $kr^*_{\rm eff}\ll{1}$, 
which is clearly seen by calculating the second order correction to the scattering amplitude.

Up to the terms $\sim(kr^*_{\rm eff})^2$, 
the on-shell scattering amplitude following form the solution of the scattering problem for particles with mass $m^{*}$, is given by~\cite{Shlyapnikov2009,Shlyapnikov2011}:
\begin{equation}\label{eq:scattS}
	f(k)=-\frac{8}{3}\frac{\hbar^2}{m}kr^{*}+\frac{\pi}{2}\frac{\hbar^2}{m}(kr^{*})^2\ln\left({Bkr^{*}}\right),
\end{equation}
where the numerical coefficient $B$ comes from short-range physics. 
For calculating $B$ we introduce a perfectly reflecting wall at intermolecular distances ${r}\sim{b}$,
which takes into account that two fermions practically can not get to one and the same lattice site. 
The coefficient $B$ depends on the ratio $r^*_{\rm eff}/b$, and we show this dependence in Fig.~4a.

\begin{figure}[h]
\begin{minipage}[h]{1\linewidth}
\centerline{\includegraphics[width=1\columnwidth]{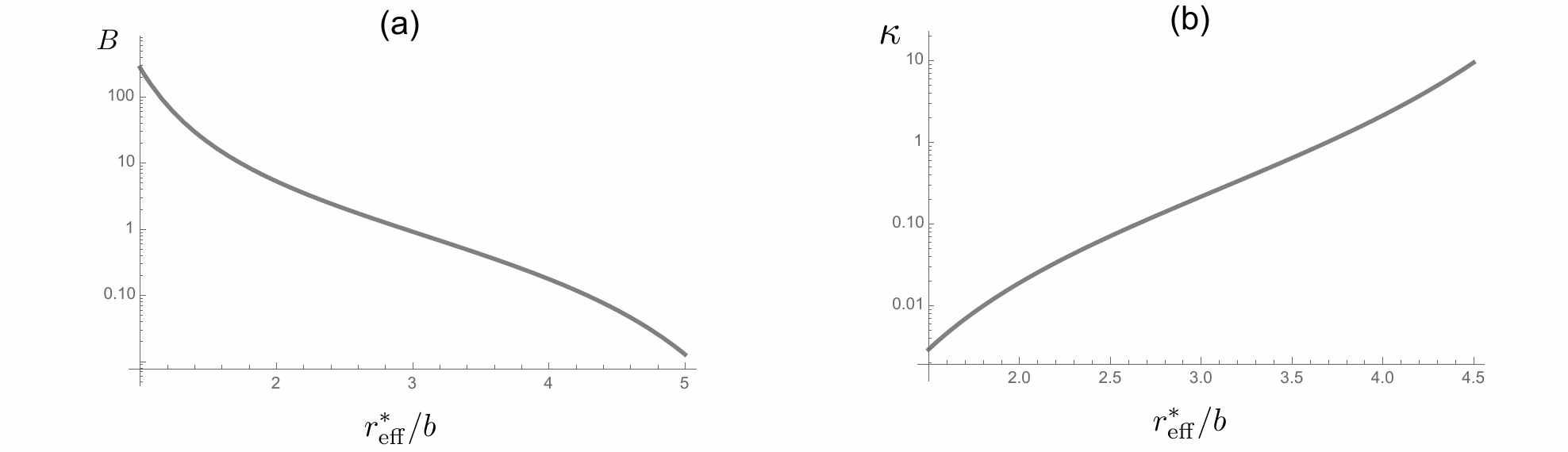}}
\centerline{\mbox{\textbf{Fig.~4}. Coefficients $B$ and $\kappa$ as functions of $r^{*}_{\rm eff}/b$.}}
\end{minipage}
\end{figure}

\subsubsection{$P$-wave pairing of microwave-dressed polar molecules in a 2D lattice}\label{sec:4pairing}

Being dressed with a microwave field, polar molecules acquire an attractive dipole-dipole tail in the interaction potential.
This leads to superfluid $p$-wave pairing of identical molecules. 
In free space the emerging ground state is the topological $p_x+ip_y$ superfluid, 
and the leading part of the scattering amplitude can be obtained in the first Born approximation~\cite{Shlyapnikov2009,Shlyapnikov2011}.   
We assume the weakly interacting regime at a small filling factor in the lattice, $k_Fb\ll{1}$.

The Hamiltonian of the system is $\hat{\mathcal{H}}=\hat{H}_{0}+\hat{H}_{\mathrm{int}}$, with
\begin{equation}\label{eq:H0lat}
	\hat{H}_{0}=\sum\nolimits_{\bf q}{\varepsilon_{\bf q}\hat{a}_{\bf q}^{\dagger}\hat{a}_{\bf q}^{}},
\end{equation}
where $\hat{a}_{\bf q}^{}$, $\hat{a}_{\bf q}^{\dagger}$ are the annihilation and creation operators of a molecule with quasimomentum ${\bf q}$, 
and $\varepsilon_{\bf q}$ is the single particle energy. 
In the low momentum limit we have $\varepsilon_{\bf q}=\hbar^2q^2/2m^{*}$,
where $m^{*}>m$ is the effective mass in the lowest Bloch band. 
The quantity $\hat{H}_{\mathrm{int}}$ describes the interaction between the molecules and is given by
\begin{equation}\label{eq:Hitnlat}
	\hat{H}_{\mathrm{int}}=
	-\frac{1}{2}\sum_{{\bf r}_j\ne{\bf r}'_j}{\hat\psi^{\dagger}({\bf r}_j)\hat\psi^{\dagger}({\bf r}'_j)\frac{d^2}{|{\bf r}_j-{\bf r}'_j|^3}\hat\psi^{}({\bf r}'_j)\hat\psi^{}({\bf r}_j)},
\end{equation}
where $\hat\psi^{}({\bf r}_j)$ is the field operator of a particle in the lattice site $j$ located at ${\bf r}_j$ in the coordinate space. 
At a small filling factor in the low momentum limit, 
the main contribution to the matrix elements of $\hat{H}_{\mathrm{int}}$ comes from intermolecular distances $|{\bf r}_j-{\bf r}'_j|\gg{b}$. 
Therefore, we may replace the summation over ${\bf r}_j$ and ${\bf r}'_j$ by the integration over $d^2{\bf r}_j$ and $d^2{\bf r}'_j$.
As a result the Hamiltonian of the system reduces to 
\begin{equation}\label{eq:H}
	\hat{H}=-\int{\frac{\hbar^2}{2m^{*}}\hat\psi^{\dagger}({\bf r})\nabla^2\hat\psi^{}({\bf r})d^2r}
	-\frac{d^2}{2}\int{\frac{d^2rd^2r\,'}{|{\bf r}-{\bf r}'|^3}\hat\psi^{\dagger}({\bf r})\hat\psi^{\dagger}({\bf r}')\hat\psi^{}({\bf r}')\hat\psi^{}({\bf r})},
\end{equation}
where the first term in the right hand side is $\hat{H}_{0}$ (\ref{eq:H0lat}) rewritten in the coordinate space.
We thus see that the problem becomes equivalent to that of particles with mass $m^{*}$ in free space.

The scattering amplitude at $k=k_F$ is obtained from the solution of the scattering problem in the lattice. 
For particles that have mass $m^*$, the amplitude is written as follows:
\begin{equation}\label{eq:scatt}
	f(k_F){=}-\frac{8}{3}\frac{\hbar^2}{m^*}k_Fr^{*}_{\rm eff}+\frac{\pi}{2}\frac{\hbar^2}{m^*}(k_Fr^{*}_{\rm eff})^2\ln\left({Bk_Fr^{*}_{\rm eff}}\right),
\end{equation}
where $k_Fr^{*}_{\rm eff}\ll1$, and $B$ is a numerical coefficient coming from short-range physics.  
Since for weak interactions two fermions practically do not get to the same lattice site, 
for calculating $B$ we may introduce a perfectly reflecting wall at intermolecular distances $r\sim b$. 
For the superfluid pairing the most important are particle momenta ${\sim}k_F$. 
Therefore, the low-momentum limit requires the inequality $k_Fb\ll 1$.
The the coefficient $B$ as the of function $r^*_{\rm eff}/b$ is shown in Fig.~4.

The treatment of the superfluid pairing is the same as in Ref.~\cite{Shlyapnikov2009}, including the Gor'kov-Melik-Barkhudarov correction~\cite{Gor'kov1961}.
We should only replace the mass $m$ with $m^{*}$. 
The expression for the critical temperature then becomes:
\begin{equation}\label{eq:TcP}
	T_c=E_F\frac{\kappa}{(k_Fr^{*}_{\rm eff})^{9\pi^2/64}}\exp\left[-\frac{3\pi}{4k_Fr^{*}_{\rm eff}}\right],
\end{equation}
where $\kappa\simeq0.19B^{-9\pi^2/64}$.
The coefficient $\kappa$ is displayed in Fig.~4b as a function of $r^*_{\rm eff}/b$.
There are two important differences of equation (\ref{eq:TcP}) from a similar equation in free space obtained in Ref.~\cite{Shlyapnikov2009}. 
First, the Fermi energy $E_F$ is smaller by a factor of $m/m^*$, and  
the effective dipole-dipole distance $r^*_{\rm eff}$ is larger than the dipole-dipole distance in free space by $m^*/m$. 
Second, the coefficient $B$ and, hence, 
$\kappa$ in free space is obtained from the solution of the Schr\"odinger equation in the full microwave-induced potential of interaction between two molecules, 
whereas here $B$ follows from the fact that the relative wavefunction is zero for $r\leq{b}$ (perfectly reflecting wall). 

It is clear that for the same 2D density $n$ (and $k_F$) the critical temperature in the lattice is larger than in free space because the BCS exponent in Eq.~(\ref{eq:TcP}) is smaller.
However, in ordinary optical lattices one has the lattice constant $b\gtrsim 200$ nm.
In this case, for $m^*/m\approx 2$ 
(still the tight binding case with $b/\xi_0\approx 3$, where $\xi_0$ is the extension of the particle wavefunction in the lattice site) 
and at a fairly small filling factor (let say, $k_Fb=0.35$)
the Fermi energy for the lightest alkaline polar molecules NaLi is about $10$ nK ($n\approx{2\times10^{7}}$ cm$^{-2}$). 
Then, for $k_Fr^*_{\rm eff}$ approaching unity the critical temperature is of the order of a nanokelvin 
(for $k_Fb=0.35$ and $r^*_{\rm eff}/b\approx{3}$ Fig.~4 gives $\kappa\sim1$).

\section{Concluding remarks}\label{sec:4conclusion} 

The critical temperature of proposed $p$-wave lattice superfluids was estimated on the level of a few nanokelvins,
which is fairly low.
The situation  is quite different in recently introduced subwavelength lattices, where the lattice constant can be as small as 
$b\simeq{50}$
nm~\cite{Zoller2008,Berman2002,Dalibard2015,Lukin2012,Cirac2013,Chen2014,Spreeuw2011,CiracKimble2015,VuleticLukin2013,KimbleChang2015,BaranovZoller2016,Gorshkov2016}.
This strongly increases all energy scales. 
Then, in the case of fermionic atoms for $k_Fb=0.5$ the density and the Fermi energy will be higher by an order of magnitude.
Hence, departing from the Feshbach resonance one gets the same BCS exponent as in the ordinary lattice,
and the critical temperature for $^{6}$Li will be $\sim$50 nK.
The recombination time is again on the level of a second.

Note that there is a (second-order) process, 
in which the interaction between two identical fermions belonging to the lowest Bloch band provides a virtual transfer of one of them to a higher band.
Then, the two fermions may get to the same lattice site and undergo the inelastic process of collisional relaxation. 
The rate constant of this second-order process is roughly equal to the rate constant in free space, 
multiplied by the ratio of the scattering amplitude (divided by the elementary cell area) to the frequency of the potential well in a given lattice site 
(the difference in the energies of the Bloch bands). 
This ratio originates from the virtual transfer of one of the fermions to a higher band and does not exceed $(\xi/b)^2$. 
Even in not a deep lattice, where $m^{*}/m$ is 2 or 3, we have $(\xi/b)^2<0.1$. 
Typical values of the rate constant of inelastic relaxation in free space are $\sim{10^{-8}{-}10^{-9}}$ cm$^2$/s~\cite{Shlyapnikov2009}, 
and hence in the lattice it will be lower than $10^{-9}$ or even $10^{-10}$~cm$^2$/s. 
Thus, the rate of this process is rather low and for densities approaching $10^{9}$~cm$^{-2}$ the decay time will be on the level of seconds or even tens of seconds.

For NaLi molecules taking $b=50$ nm and $k_Fb=0.35$,
the Fermi energy exceeds 200 nK ($n\approx{4\times10^{8}}$ cm$^{-2}$).
Then, for the same $\kappa\sim1$ and $k_Fr^*_{\rm eff}$ approaching unity we have $T_c\sim{20}$ nK, which is twice as high as in free space. 
An additional advantage of the lattice system is the foreseen quantum information processing, since addressing qubits in the lattice is much easier than in free space.  

Superfluidity itself can be detected in the same way as in the case of $s$-wave superfluids~\cite{Giorgini2008,Zwierlein2005}. 
Rotating the $p_x+ip_y$ superfluid and inducing the appearance of vortices 
one can find signatures of Majorana modes on the vortex cores in the RF absorption spectrum~\cite{Cooper2007}.
Eventually, one can think of revealing the structure of the order parameter by visualizing vortex-related dips in the density profile
on the approach to the strongly interacting regime,
where these dips should be pronounced at least in time-of-flight experiments. 

\section*{Acknowledgments} 

The research leading to these results has received funding from the European Research Council under European Community's Seventh Framework Programme 
(FP7/2007-2013 Grant Agreement no. 341197).
The work was carried out with financial support from the Ministry of Education and Science of the Russian Federation in the framework of increase Competitiveness Program of NUST ''MISIS", 
implemented by a governmental decree dated 16th of March 2013, No 211.
The work was also supported in part by the RFBR Grant 17-08-00742 (A.F.).

\end{document}